\documentclass[onecolumn,preprintnumbers,amsmath,amssymb,superscriptaddress]{revtex4}

\usepackage{graphicx}% Include figure files
\usepackage{dcolumn}% Align table columns on decimal point
\usepackage{bm}% bold math
\usepackage{longtable}% bold math

\usepackage{color}
\usepackage{latexsym}
\usepackage{amssymb}
\usepackage{amsfonts}
\usepackage{amsmath}

\usepackage[latin1]{inputenc}
\usepackage[english]{babel}

\def\be{\begin{equation}}
\def\ee{\end{equation}}
\def\bc{\begin{center}}
\def\ec{\end{center}}

\def\be{\begin{equation}}
\def\ee{\end{equation}}
\def\bc{\begin{center}}
\def\ec{\end{center}}

\begin{document}

\title{Acquaintance role for decision making and exchanges\\ in social networks}

\author{Elena Agliari}
\affiliation{Dipartimento di Fisica, Universit\`a di Parma, Italy}
\affiliation{INFN, Gruppo Collegato di Parma, Italy}
\author{Adriano Barra}
\affiliation{Dipartimento di Fisica, Sapienza Universit\`a di Roma, Italy}
\author{Raffaella Burioni}
\affiliation{Dipartimento di Fisica, Universit\`a di Parma, Italy}
\affiliation{INFN, Gruppo Collegato di Parma, Italy}
\author{Pierluigi Contucci} 
\affiliation{Dipartimento di Matematica, Universit\`a di Bologna, Italy}

\begin{abstract}
We model a social network by a random graph whose nodes
represent agents and links between two of them stand for a reciprocal
interaction; each agent is also associated to a binary variable which represents a dichotomic opinion or attribute. We consider both the case of pair-wise ($p=2$) and multiple ($p>2$) interactions among agents and we study the behavior of the resulting system by means of the
energy-entropy scheme, typical of statistical mechanics
methods.
We show, analytically and numerically, that the
connectivity of the social network plays a non-trivial role: while for pair-wise
interactions ($p=2$) the connectivity weights linearly, when interactions
involve contemporary a number of agents larger than two ($p>2$),
its weight gets more and more important. As a result, when $p$ is
large, a full consensus within the system, can be reached at relatively small critical couplings with
respect to the $p=2$ case usually analyzed, or, otherwise stated,
relatively small coupling strengths among agents are sufficient to orient
most of the system.

\end{abstract}

%\author{Elena Agliari\footnote{Dipartimento di Fisica, Universit\`{a} di Parma},
%Adriano Barra\footnote{Dipartimento di Fisica, Sapienza
%Universit\`{a} di Roma}, Raffaella Burioni$^*$, Pierluigi
%Contucci\footnote{Dipartimento di Matematica, Universit\`{a} di
%Bologna}}

\maketitle

\section{Introduction}
In recent years there has been an increasing interest in exploring
a quantitative approach to social sciences by means of statistical
physics methods \cite{ising,bovier,pc2,pc0,pc1,galam}. In particular, a great attention
has been paid to the topological structure representing the network of interaction
between agents, which is known to deeply influence the global behavior. 
Social systems are effectively envisaged by graphs,
whose nodes represent agents and links between them
represent a couple interaction, and  if relations between multiple agents are
present, the network is represented by an hypergraph, formed by n-simplex
interactions \cite{zicco}.  In this framework, the behavior of a large
number of interacting units, i.e. a social group, can be investigated by means of the
energy-entropy scheme (free-energy variational principle), typical of statistical mechanics
methods.

The description of both the network topology, i.e. who is
interacting with whom, and of the coupling strength between agents, i.e.
the magnitude of each link, can be attained via a cost function
$H$. In particular, we choose a class of such functions which may
mimic several different contexts among $N$ agents: Each cost
function displays the mean-field form $\sum^N_{{i_1} < {i_2} < ...
< {i_p}} J_{i_1 \, i_2 \, ... i_p} \sigma_{i_1} \times
\sigma_{i_2} \times ... \times \sigma_{i_p}$, where $\sigma_i =
\pm 1$, and the tensor $\mathbf{J}$ tunes the interaction
strength in the network. Restricting for simplicity to the cases
$p=2,3$, we deal with agents interacting in couples and triplets
(strictly speaking, for $p=2$ we talk about networks and for $p>2$ we talk
about hypergraphs \cite{zicco}).

As for the network topology, many structures have been proposed
along recent years, and they are in general built starting from three main
architectures: the random graph, the small-world and the
scale-free network \cite{caldarelli}, which reproduce some general features
of the observed social systems; here we focus on the former
as it allows a simpler mathematical approach.

A social system with this simple formal description can be used
to model decision making:
$\sigma_i$ is the opinion of the agent $i$ and each agent $i$
tries to aline his opinion (in the case of imitative behavior, i.e. $J_{ij} >0$) to other agents' 
viewpoint, interacting one by one ($p=2$), or in larger groups (i.e. $p \geq 3$).

Another appealing application of this models concerns trading among agents:
Suppose we represent a market society only with couple exchanges
$(p=2)$. Then, there are just sellers and buyers and they interact
only pairwise. In this case if the buyer $i$ has money
($\sigma_i=+1$) and the seller $j$ has the product
($\sigma_j=+1$), or if the buyer has no money and the seller has
no products ($\sigma_{i}=\sigma_{j}=-1$), the two merge their will
and the cost function reaches the minimum. Otherwise, if the
seller has the product but the buyer has no money (or viceversa),
their two states are different ($\sigma_{i} \sigma_{j}=-1$) and
the cost function is not minimized. In this scenario, the
possibility that an agent is satisfied increases only linearly
with the number of his/her acquaintances, namely with the degree
of the relevant node. In fact, the higher the number of
``neighbors", the larger the possibility of trading.

When switching to the case $p=3$, other strategies (on the
timescale by which the connectivity remains constant) are
available: for example the buyer may not have the money, but he may have a valuable good
which can be offered to a third agent, who takes it and, in change, gives to the seller the money,
so that the buyer can obtain his target by using a barter-like approach.
In this case the contribution of the third agent $k$ can either
avoid the two frustrated configurations of the previous picture,
by providing a factor $\sigma_k=-1$, or it can leave the
frustration unaffected if he does not agree $(\sigma_k=+1)$.
Interestingly, we find that in this case $(p=3)$, the amount of
acquaintances one is in touch with (strictly speaking, the degree
of connectivity $\alpha$) does not contribute linearly as for
$p=2$, but quadratically: this seems to suggest that if a society
deals primarily with direct exchanges, no particular effort should
be done to connect people, while, if barter-like approaches are
allowed, then the more connected the society, the larger the
satisfaction reached on average by each agent in his specific
goal. Intuitively, the above scenario seems to match the contrast
among the classical barter-like approach of villages, where,
thanks to the small amount of citizens, their degree of reciprocal
knowledge is quite high and the money-mediated one of citizens in
big metropolis, where a real reciprocal knowledge is fewer.

In this work we want to pave both the analytical and the numerical analysis
of what settled so far: to tackle this task, at first we build our cost
function in the next section $2$, then in
section $3$ we analyze it by equilibrium statistical mechanics of
disordered systems and in section $4$ we corroborate our findings by
Monte Carlo numerical simulations.

%$$
%H(s)= -J \sum_{i_1<i_2<...<i_p}^ N s_{i_1} \times s_{i_2} \times
%... \times s_{i_p}
%$$

\section{Definition of the model}
First of all, we define a suitable
Hamiltonian (a cost function) acting on a random network with
connectivity $\alpha$ made up of $N$ agents $\sigma_{i}=\pm1, \ i \in
[1,N]$.
\newline
Introducing $p$ families of i.i.d. random variables
$\{i_{\nu}^1\},\{i_{\nu}^2\},...,\{i_{\nu}^p\}$ uniformly
distributed on the previous interval, the Hamiltonian is given by
the following expression
\begin{equation}
H_{N}(\sigma,\gamma(\alpha))=- J \sum_{\nu=1}^{k_{\gamma (\alpha)
N}} \sigma_{i_{\nu}^1}\sigma_{i_{\nu}^2}...\sigma_{i_{\nu}^p}
\end{equation}
where, reflecting the underlying Erd\"{o}s-Renyi graph, $k$ is a
Poisson distributed random variable with mean value
$\gamma(\alpha) N$ and $J>0$ is the interaction strength, supposed to be
the same for each $p$-plet. The relation among the coordination
number $\alpha$ and $\gamma$ is $\gamma \sim \alpha^{p-1}$: this
will be easily understood a few lines later by a normalization
argument coupled with the high connectivity limit of this mean
field model.

The quenched expectation of the model is given by the composition
of the Poissonian average with the one performed over the families
$\{i_{\nu}\}$
\begin{equation}
\textbf{E}[\cdot] = E_PE_i[\cdot] = \sum_{k=0}^{\infty}
\frac{e^{-\gamma(\alpha) N}(\gamma(\alpha) N)^k}{k!} \times
\frac{1}{N^p}\sum_{i_{\nu}^1....i_{\nu}^p}^{1,N}[\cdot].
\end{equation}
Following a statistical mechanics (SM) approach, we know that the
macroscopic behavior of the system as a function of the average
degree $\alpha$ and of the interaction strength $J$, is described
by the following free energy density
\begin{equation}
A(\alpha,J)= \lim_{N \to \infty} A_N(\alpha,J) = \lim_{N \to
\infty}\frac1N\textbf{E}\ln\sum_{\sigma}\exp(-
H_{N}(\sigma,\gamma(\alpha);J)).\
\end{equation}
The normalization constant can be extracted performing the expectation
value of the cost function:
\begin{equation}
\textbf{E}[H] = - \sum_{k=0}^{\infty} \frac{e^{-\gamma N}(\gamma
N)^k}{k!} \times \frac{J}{N^p}\sum_{i_{\nu}^1....i_{\nu}^p}^{1,N}
\sum_{\nu=1}^{k_{\gamma N}}
\sigma_{i_{\nu}^1}\sigma_{i_{\nu}^2}...\sigma_{i_{\nu}^p} =
-\gamma J N m^p,
\end{equation}
by which it is easy to see that the model is well defined and, in
particular, it is linearly extensive in the volume. Then, in the
high connectivity limit, each agent interacts with all the others
and, in the thermodynamic limit, the coordination number $\alpha
\to \infty$ as $N$. Now, if $p=2$ the amount of couples in the
summation scales as $N(N-1)/2$ and $\gamma = 2\alpha$ provides the
right scaling; if $p=3$ the amount of triples scales as
$N(N-1)(N-2)/3!$ and $\gamma = 3!\alpha^2$ again recovers the
right connectivity behavior. The result ca be generalized to every finite
$p<N$.

Finally, we introduce the fundamental quantities expressed by the
multi-overlap \be
q_{1...n}=\frac1N\sum_{i=1}^{N}\sigma^{(1)}_{i}...\sigma^{(n)}_{i}
, \ee with a particular attention to the magnetization $m = q_1
=(1/N)\sum_{i=1}^{N}\sigma_{i}$. This plays the role of order
parameter, representing the average opinion in decision making and
the average trade in market.

\section{Analytical results}

In this section we summarize the scheme developed in statistical
mechanics: Our goal is finding an explicit expression for the
minimized free energy, which describes the overall behavior of our
agents. To this task we decompose this quantity via the next
equation (\ref{main}) (whose proof is known in SM
\cite{abarra,barra}) into two quantities which can be estimated in
an easier way, namely a cavity function $\Psi$ and a connectivity
shift $d_{\alpha}A$: \be\label{main} A(\alpha,J) = \ln2
-\frac{\alpha}{p-1}\frac{d}{d\alpha}A(\alpha,J) + \Psi(\alpha,J),
\ee where the cavity function $\Psi(\alpha,J) =
\lim_{N\rightarrow\infty}\Psi_N(\gamma,J)$ is defined, at finite
$N$, as \be \Psi_N(\gamma,J) =
\textbf{E}\Big[\ln\frac{\sum_{\{\sigma\}}
e^{J\sum_{\nu=1}^{k_{\gamma N}}
\sigma_{i_{\nu}^1}\sigma_{j_{\nu}^2}...\sigma_{j_{\nu}^p}}\;
e^{J\sum_{\nu=1}^{k_{2\gamma }}
\sigma_{i_{\nu}^1}\sigma_{j_{\nu}^2}...\sigma_{j_{\nu}^{p-1}}}}
{\sum_{\{\sigma\}} e^{J\sum_{\nu=1}^{k_{\gamma N}}
\sigma_{i_{\nu}^1}\sigma_{j_{\nu}^2}...\sigma_{j_{\nu}^p}}}\Big] =
\textbf{E}\Big[\ln \frac{\sum_{\{\sigma\}}e^{-
H_{N+1}(\sigma,\gamma; J)}} {\sum_{\{\sigma\}}e^{-
H_N(\sigma,\gamma; J)}}\Big]. \nonumber \ee Hence, we now need to
evaluate the cavity function and the connectivity shift (the
$\alpha$ derivative of the free energy density). Starting with the
latter and using the following properties of the Poisson
distribution \be \textbf{E}[kg(k)] = N\gamma\textbf{E}[g(k+1)], \
\ \frac{d}{d\gamma}\textbf{E}[g(k)] =
 N\textbf{E}[g(k+1) - g(k)],
\ee we can write
\begin{eqnarray}
\frac{d}{d\alpha}A(\alpha,J) &=&
\frac{1}{N}\frac{d}{d\alpha}\textbf{E}\Big[\ln
\sum_{\{\sigma\}}e^{- H_N(\sigma,\gamma;J)}\Big]
= \frac{(p-1)}{N}\alpha^{p-2}\frac{d}{d\gamma}\textbf{E}\Big[\ln \sum_{\{\sigma\}}e^{- H_N(\sigma,\gamma;J)}\Big]  \nonumber \\
&=& (p-1)\alpha^{p-2}\textbf{E}\Big[\ln \sum_{\{\sigma\}}
e^{J\sum_{\nu=1}^{k+1} \sigma_{i_{\nu}^1}...\sigma_{i_{\nu}^p}} -
\ln \sum_{\{\sigma\}} e^{J\sum_{\nu=1}^{k}
\sigma_{i_{\nu}^1}...\sigma_{i_{\nu}^p}}\Big]. \nonumber
\end{eqnarray}
Now, considering the following relation and definition
\begin{eqnarray}
e^{J\sigma_{i^1}...\sigma_{i^p}} = \cosh J +
\sigma_{i^1}...\sigma_{i^p}\sinh J,  \ \ \theta \equiv \tanh J,
\end{eqnarray}
and expanding the logarithm, we obtain
\begin{equation}
\frac{d}{d\alpha}A(\alpha,J) = (p-1)\alpha^{p-2}\ln\cosh J -
(p-1)\alpha^{p-2}\sum_{n=1}^{\infty}\frac{(-1)^n}{n}\theta^n
\langle q_{1,...,n}^p \rangle.
\end{equation}
With the same procedure, posing $\beta' \equiv 2\alpha\theta(J)$,
and using a little of algebra \cite{full,BCC}, it is possible to
show that \be \Psi_N(\alpha,J) = 2\alpha^{p-1}\ln\cosh J +
\frac{\beta'^2}{2}\langle m_1^{2(p-1)}\rangle  -
\frac{\beta'^2\theta^2}{4}\langle q_{12}^{2(p-1)}\rangle  +
O(\tilde{\beta}^3) \nonumber \ee and we get the result: a
polynomial form for the free energy density \be A(\alpha,J)= \ln2
+ \alpha^{p-1}\ln\cosh J + \frac{\beta'}{2}\Big(\beta'\langle
m^{2(p-1)}\rangle - \langle m^{p}\rangle\Big) + O(\beta'^3).
\nonumber \ee It is straightforward to see that for $p=2$ the well
known diluted Curie-Weiss is recovered as well as its criticality
at $\beta'=1$ \cite{ABC}. Further, it is enough to explore the
ergodic phase ($m \equiv 0$) to see that $\alpha$ appears at the
power $p-1$ (i.e. $2$ for $p=3$).

\section{Numerical results}
We now analyze the system introduced in Sec.~$2$, from the numerical point of view by performing extensive Monte Carlo simulations \cite{barkema}; here we especially focus on the cases $p=2$ and $p=3$.
\begin{figure}[tb]
\begin{center}
\includegraphics[width=60mm, height=50mm]{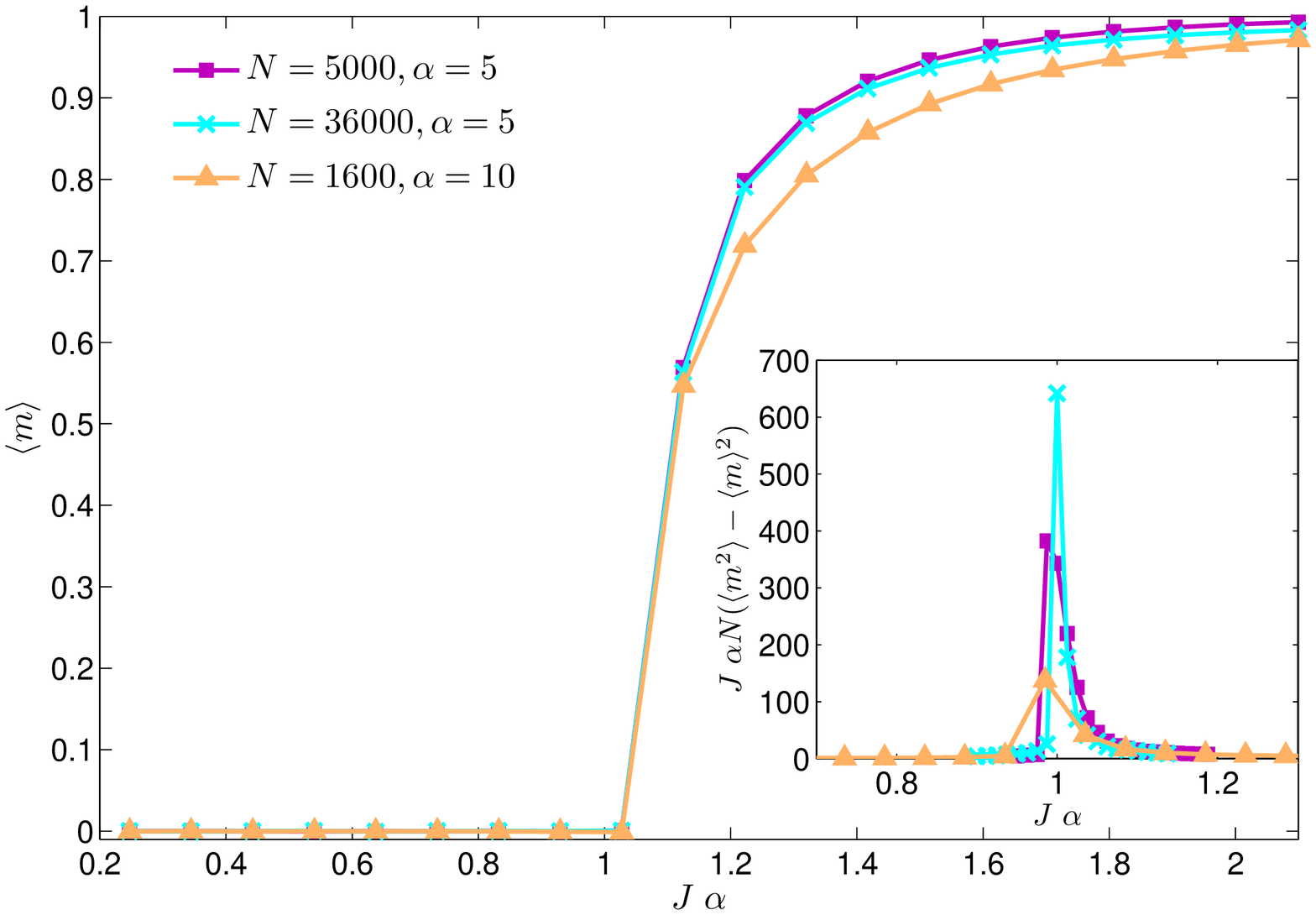}
\includegraphics[width=60mm, height=50mm]{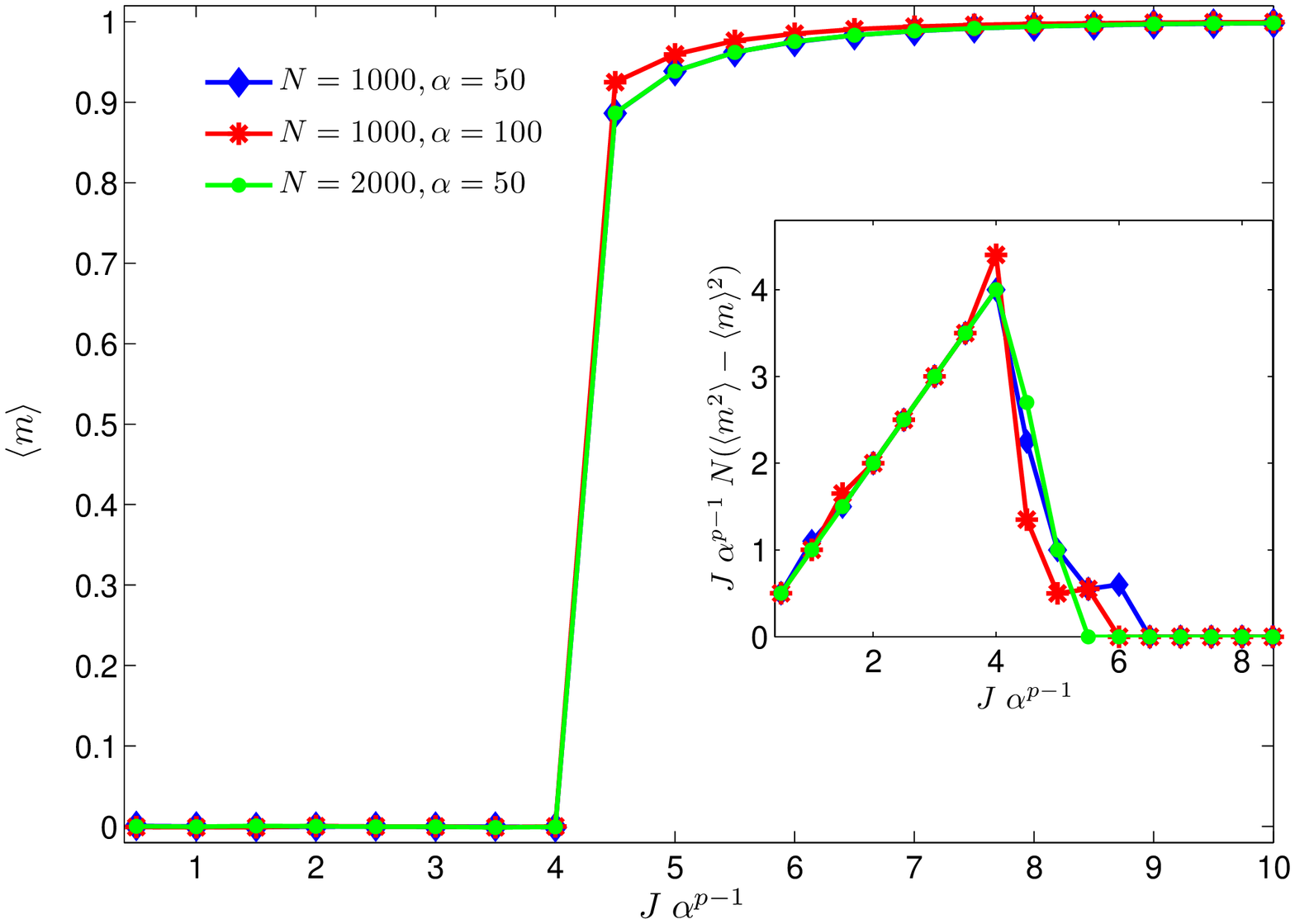}
\caption {Magnetization (main figure) and its normalized
fluctuations (inset) for systems of different sizes and different
dilution as a function of $J \; \alpha^{p-1}$. Left panel: $p=2$,
right panel $p=3$. The collapse of all the curves provides a
strong evidence for the scaling of the interaction strength,
coupled to the connectivity.} \ec
\end{figure}
In general, we find that the system is able to relax to a
well-defined steady state characterized by average observables
(such as total energy and magnetization) which are independent of
the initial configuration and of the system size $N$, as long as
$N$ is large enough to avoid finite-size effects. On the other
hand, the average observables vary as the average coordination
number $\alpha$ and/or the interaction strength $J$ are tuned. In
particular, as shown in Fig.~$1$ we find that the curves for the
order parameter collapse when plotted as a function of $J \;
\alpha^{p-1}$, confirming the scaling found analytically. We also
notice that the system exhibits a phase transition at a critical
interaction strength $J_c$, which depends on the connectivity of
the underlying network according to $J_c = \alpha^{-1}$ for $p=2$
and $J_c = 4\alpha^{-2}$ for $p=3$. The existence of a phase
transition is also confirmed by the corresponding peak displayed
by the fluctuations of the order parameter $\langle m^2 \rangle -
\langle m \rangle^2$, see insets in Fig.~$1$.

Before concluding, we stress the non-trivial role played by the
connectivity of the social network: while for pair-wise
interactions ($p=2$) $\alpha$ weights linearly, when interactions
involve contemporary a number of agents larger than two ($p>2$),
its weight gets more and more important. As a result, when $p$ is
large, a full consensus within the system, i.e. a ferromagnetic
state, can be reached at relatively small critical strengths with
respect to the $p=2$ case usually analyzed, or, otherwise stated,
relatively small coupling strengths among agents are sufficient to orient
most part on the social system \cite{ERP}.

\end{document}